\documentclass{article}

\PassOptionsToPackage{numbers, compress}{natbib}

\usepackage[preprint]{neurips_2019}

\usepackage[utf8]{inputenc} 
\usepackage[T1]{fontenc}    
\usepackage{hyperref}       
\usepackage{url}            
\usepackage{booktabs}       
\usepackage{amsfonts}       
\usepackage{nicefrac}       
\usepackage{microtype}      
\usepackage{graphicx}
\usepackage{subfigure}
\usepackage{wrapfig}
\usepackage{amsmath}
\usepackage{amssymb}
\usepackage{amsthm}
\usepackage{bbm}
\usepackage{dsfont}
\usepackage{makecell}
\usepackage{paralist}
\usepackage{multirow}
\usepackage{color}

\def\a{{\boldsymbol a}}
\def\B{{\boldsymbol B}}
\def\b{{\boldsymbol b}}

\def\h{{\boldsymbol h}}
\def\H{{\boldsymbol H}}
\def\G{{\boldsymbol G}}
\def\N{{\boldsymbol N}}

\def\X{{\boldsymbol X}}
\def\Y{{\boldsymbol Y}}
\def\y{{\boldsymbol y}}

\def\S{{\boldsymbol S}}
\def\x{{\boldsymbol x}}
\def\y{{\boldsymbol y}}
\def\z{{\boldsymbol z}}
\def\Z{{\boldsymbol Z}}
\def\M{{\boldsymbol M}}

\def\U{{\boldsymbol U}}
\def\u{{\boldsymbol u}}
\def\V{{\boldsymbol V}}
\def\v{{\boldsymbol v}}
\def\W{{\boldsymbol W}}
\def\w{{\boldsymbol w}}
\def\X{{\boldsymbol X}}
\def\z{{\boldsymbol z}}
\def\Z{{\boldsymbol Z}}
\def\0{{\boldsymbol 0}}
\def\1{{\boldsymbol 1}}

\def\OM{{\mathcal O}}

\def\JM{{\mathcal J}}

\def\RB{{\mathbb R}}

\def\LM{{\mathcal L}}

\def\tanh{\text{tanh}}
\def\sgn{\text{sign}}

\def\1{\mathds{1}}
\def\bone{{\bf 1}}
\def\bzero{{\bf 0}}

\title{On the Evaluation Metric for Hashing}

\author{%
  Qing-Yuan Jiang, Ming-Wei Li {\normalfont and} Wu-Jun Li\\
  National Key Laboratory for Novel Software Technology\\
  Collaborative Innovation Center of Novel Software Technology and Industrialization\\
  Department of Computer Science and Technology, Nanjing University, China\\
  \texttt{qyjiang24@gmail.com, limw@lamda.nju.edu.cn, liwujun@nju.edu.cn} \\
}

\begin{document}

\maketitle
\begin{abstract}
 Due to its low storage cost and fast query speed, hashing has been widely used for large-scale approximate nearest neighbor~(ANN) search. Bucket search, also called hash lookup, can achieve fast query speed with a sub-linear time cost based on the inverted index table constructed from hash codes. Many metrics have been adopted to evaluate hashing algorithms. However, all existing metrics are improper to evaluate the hash codes for bucket search. On one hand, all existing metrics ignore the retrieval time cost which is an important factor reflecting the performance of search. On the other hand, some of them, such as mean average precision~(MAP), suffer from the uncertainty problem as the ranked list is based on \emph{integer-valued} Hamming distance, and are insensitive to Hamming radius as these metrics only depend on \emph{relative} Hamming distance. Other metrics, such as precision at Hamming radius $R$, fail to evaluate global performance as these metrics only depend on one specific Hamming radius. In this paper, we first point out the problems of existing metrics which have been ignored by the hashing community, and then propose a novel evaluation metric called \underline{r}adius \underline{a}ware \underline{m}ean \underline{a}verage \underline{p}recision~(RAMAP) to evaluate hash codes for bucket search. Furthermore, two coding strategies are also proposed to qualitatively show the problems of existing metrics. Experiments demonstrate that our proposed RAMAP can provide more proper evaluation than existing metrics.
\end{abstract}

\section{Introduction}
Approximate nearest neighbor~(ANN)~\cite{DBLP:conf/vldb/GionisIM99,DBLP:conf/compgeom/DatarIIM04,DBLP:conf/focs/AndoniI06} search plays a fundamental role in a wide range of areas, including machine learning~\cite{DBLP:conf/icml/LiuWKC11,DBLP:conf/kdd/YangZZX019,DBLP:journals/tkdd/YangWSLZXY22,DBLP:journals/datamine/LiYZ23,DBLP:conf/aaai/YangHGXX23,DBLP:journals/tois/YangZSDZL24}, data mining~\cite{DBLP:conf/kdd/Li15}, multimodal learning~\cite{DBLP:journals/pami/TangSQLWYJ17,DBLP:conf/mm/FuJQSJCH18,DBLP:conf/cvpr/TomeiCBC19,DBLP:conf/icdm/RazzakYYX19,DBLP:journals/tkde/YangFZLJ21,DBLP:journals/tkde/YangZWLXJ21,DBLP:conf/mm/0074ZGGZ22,DBLP:journals/titb/ZhangYYGLZYR22,DBLP:journals/toc/YangWZYXY22,DBLP:journals/tkde/YangYBZZGXY23,DBLP:journals/chinaf/YangBGZYY23,DBLP:journals/tkde/YangYBZZGXY23,DBLP:journals/chinaf/YangBGZYY23,DBLP:journals/tgrs/MengWMYX24}, and information retrieval~\cite{DBLP:conf/sigir/ZhangWCL10,DBLP:conf/ijcai/YangZXYZY21,DBLP:conf/icme/WanWGY24,DBLP:journals/fcsc/YangGLLLY24}, and so on. As a popular technique of ANN search, hashing~\cite{DBLP:conf/nips/WeissTF08,DBLP:journals/ijar/SalakhutdinovH09,DBLP:conf/sigir/ZhangWCL10,DBLP:conf/icml/LiuWKC11,DBLP:conf/nips/KongL12,DBLP:conf/nips/Shrivastava014,DBLP:conf/icml/YuKGC14,DBLP:conf/cvpr/ShenSLS15,DBLP:conf/nips/LiSHT17,DBLP:conf/icassp/SablayrollesDUJ17,DBLP:conf/icml/DaiGKHS17,DBLP:conf/nips/SuZHT18,DBLP:conf/cvpr/0003CBS18} has attracted much attention in recent years, as it can enable significant efficiency gains in both storage and speed.

The goal of hashing is to represent the data points as compact binary hash codes~\cite{DBLP:conf/icml/NorouziF11,DBLP:conf/cvpr/GongL11,DBLP:conf/icml/LiLSHD13} which can preserve the similarity in the original space. On one hand, the storage cost will be dramatically reduced by representing data as hash codes. On the other hand, based on hash codes, fast query speed can be achieved. Specifically, there are two widely used procedures to perform hash codes based search, i.e., Hamming ranking and hash lookup~\cite{DBLP:conf/nips/LiSHT17,DBLP:conf/icml/LiuWKC11,DBLP:conf/cvpr/GongL11,DBLP:conf/cvpr/LiuWJJC12,DBLP:conf/cvpr/ShenSLS15}. The Hamming ranking procedure tries to utilize Hamming distance between query and database points to obtain a ranked list. During this procedure, one can improve the query speed by utilizing bit-wise operation to compute the Hamming distance and an $\OM(N)$ ranking algorithm to generate a ranked list, where $N$ is the number of data points in the database. Hash lookup, also called bucket search, reorganizes hash codes as an inverted index table, based on which fast query speed with a sub-linear time cost can be achieved. In practice, hash lookup is more practical than Hamming ranking for fast search, especially for cases with large-scale datasets.

Over the past decades, many hashing methods have been proposed to improve retrieval accuracy. To evaluate these methods, many metrics, such as mean average precision~(MAP), precision and recall at Hamming radius $R$, are used to evaluate the hash codes generated by hashing methods. MAP tries to evaluate the ranked list by averaging the precision at each position in the ranked list which is generated according to Hamming ranking. Precision and recall at radius $R$ aim to calculate the accuracy of the returned points whose Hamming distance to the query is less than or equals to $R$. Almost all hashing algorithms~\cite{DBLP:conf/nips/LiuMKC14,DBLP:conf/nips/Raziperchikolaei16,DBLP:conf/cvpr/Liu0SC16,DBLP:conf/nips/SuZHT18} utilize part or all of the above three metrics for evaluation.

However, we find that the above metrics have some problems. Firstly, all of them ignore the retrieval time cost which is an important factor reflecting the performance of search and should not be ignored. Secondly, MAP suffers from an uncertainty problem as the ranked list is based on \emph{integer-valued} Hamming distance~\cite{DBLP:conf/cvpr/0003CBS18}. That is to say, there might exist different MAP values for the same set of hash codes for a dataset. Furthermore, MAP is not sensitive to Hamming radius because MAP only depends on \emph{relative} Hamming distance. Thirdly, precision and recall at radius $R$ cannot evaluate global performance because these metrics only depend on one specific Hamming radius. Hence, when we use these existing metrics to evaluate two hashing algorithms, it might be difficult to decide which algorithm is better.

In this paper, we focus on the evaluation metric for hashing, and try to solve the problems mentioned above. The contributions of this paper are listed as follows:
\begin{inparaenum}[a)]
\item We point out the problems of existing metrics which have been ignored by the hashing community. To the best of our knowledge, this is the first work to systematically analyze the problems of existing evaluation metrics for hashing.
\item We propose a novel evaluation metric, called \underline{r}adius \underline{a}ware \underline{m}ean \underline{a}verage \underline{p}recision~(RAMAP), to evaluate hash codes for bucket search.
\item We propose two coding strategies to qualitatively show the problems of existing evaluation metrics.
\item Experimental results demonstrate that our proposed metric RAMAP can provide more proper evaluation than existing metrics.
\end{inparaenum}

\section{Preliminaries}
\subsection{Notation}
In this paper, we utilize boldface lowercase letters like $\w$ to denote vectors and boldface uppercase letters like $\W$ to denote matrices. The element at position $(i,j)$ of $\M$ is denoted as $M_{ij}$. We use $\bone$ and $\bzero$ to denote a vector with all elements being 1 and 0, respectively. $C_{r}^{t}$ denotes the combinatorial number of ways to pick $t$ unordered outcomes from $r$ possibilities, i.e., $C_r^t=\frac{r!}{t!(r-t)!}$. Furthermore, $\1(\cdot)$ is used to denote an indicator function where $\1(True)=1$ and $\1(False)=0$. $\sgn(\cdot)$ is an element-wise sign function where $\sgn(x)=1$ if $x\ge 0$ else $\sgn(x)=-1$.

Assume that we have $N$ database data points $\X=\{\x_i\}_{i=1}^N$ and $M$ query data points $\Y=\{\y_j\}_{j=1}^M$. We use $\G\in\{0,1\}^{N\times M}$ to denote if a data point $\x_i$ is a ground-truth neighbor of query $\y_j$. If $\x_i$ is a ground-truth neighbor of $\y_j$, $G_{ij}=1$, else $G_{ij}=0$. We use $\U=[\u_1,\dots,\u_N]^T\in\{0,1\}^{N\times Q}$ and $\V=[\v_1,\dots,\v_M]^T\in\{0,1\}^{M\times Q}$ to denote the $Q$-bits hash code for database points and query points, respectively. $\text{dist}_H(\cdot,\cdot)$ is used to denote the Hamming distance between two hash codes. For learning based hashing~\cite{DBLP:journals/ijar/SalakhutdinovH09,DBLP:conf/cvpr/GongL11}, we utilize $\Z=\{\z_k\}_{k=1}^L$ to denote training set. In practice, training set is usually sampled from the database set, i.e., $\Z\subseteq\X$. We utilize $\H=[\h_1,\dots,\h_L]^T\in\{0,1\}^{L\times Q}$ to denote the binary hash codes for training set. Furthermore, for supervised hashing, the similarity $\S\in\{0,1\}^{L\times L}$ is also available during training. If $\z_k$ and $\z_l$ are similar, $S_{kl}=1$, else $S_{kl}=0$.

Given a query hash code $\v_j$, we utilize $\b_{j_{l}}^{(R)}$ to denote the $j_l$-th bin where $R$ denotes the Hamming distance between the query and the data points in this bin. We define $r$-Hamming ball as $\B_j^{(r)}\doteq\{\x_i\;\vert\;\forall \x_i\in\X,\text{dist}_H(\u_i,\v_j)=r\}$ and set of ground-truth neighbors as $\N_j\doteq\{\x_i\;\vert\;\forall \x_i\in\X,G_{ij}=1\}$. Then we set $N^{+}_{j}=\vert\N_j\vert$, $N_{j,r}=\vert\B_j^{(r)}\vert$ and $N^{+}_{j,r}=\vert\N_j\cap\B_j^{(r)}\vert$.

\subsection{Hash Codes based Retrieval}
\begin{wrapfigure}{r}{0.5\textwidth}\vspace{-18pt}
\centering
\includegraphics[width=0.5\textwidth]{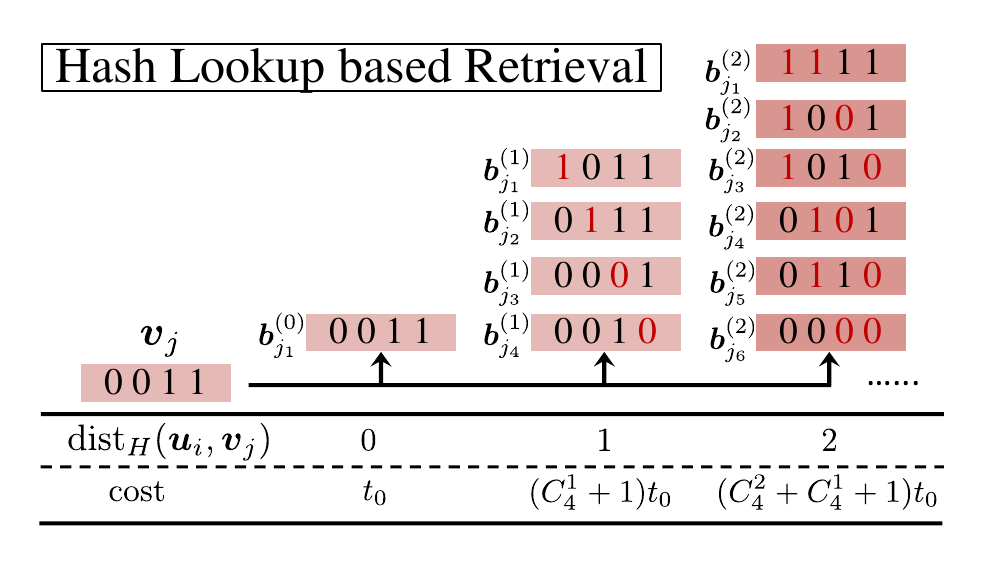}\vspace{-15pt}
\caption{An example of hash lookup.}\vspace{-20pt}
\label{fig:HL}
\end{wrapfigure}

Hamming ranking and hash lookup are two important procedures for hash codes based retrieval. Compared with Hamming ranking procedure, hash lookup procedure can achieve sub-linear query speed. Hence it is more practical in real applications. The hash lookup procedure is shown in Figure~\ref{fig:HL}. In Figure~\ref{fig:HL}, the hash code of the query is denoted as $\v_j=[0,0,1,1]^T$, the data points in the database is reorganized as an inverted index table constructed from hash codes.

Given a query with hash code $\v_j=[0,0,1,1]^T$, hash lookup~(bucket search) tries to retrieve enough candidates from the nearest tables. That is to say, this procedure will retrieve the bins $\{\b_{j_1}^{(0)};\b_{j_1}^{(1)},\b_{j_2}^{(1)}\dots\}$ sequentially until the enough candidates are gathered. Assume that the time cost for one bucket search operation is $t_0$, Figure~\ref{fig:HL} presents the time costs when we increase the query Hamming radius.

\subsection{Mean Average Precision~(MAP)}
MAP is a widely used metric for hashing~\cite{DBLP:conf/nips/LiSHT17,DBLP:conf/nips/SuZHT18}. The core idea of MAP is to evaluate a ranked list by averaging the precision at each position. Given $M$ queries $\{\y_j\}_{j=1}^M$, MAP is calculated as follows:
\begin{align}
\text{AP}(\y_j)=\frac{1}{N^{+}_j}\sum_{k=1}^N\text{precision}(k)\1(G_{kj}=1),\quad\text{MAP}=\frac{1}{M}\sum_{j=1}^M\text{AP}(\y_j).\nonumber
\end{align}
where $\text{precision}(k)$ denotes the precision at cut-off $k$ in the ranked list. When we utilize MAP to evaluate a hashing algorithm, the Hamming distance between query and database is used to obtain the ranked list.

\subsection{Precision and Recall}
Other important metrics are precision and recall at Hamming radius $R$~\cite{DBLP:conf/cvpr/LiuWJJC12,DBLP:conf/nips/LiuMKC14}. The core idea of precision@$R$~(recall@$R$) is to calculate the accuracy of returned samples where the Hamming distance between query and samples is less than or equals to $R$. Given $M$ queries $\{\y_j\}_{j=1}^M$, the precision and recall at $R$ can be calculated as follows:
\begin{align}
\text{precision}@R=\frac{1}{M}\sum_{j=1}^M\frac{\sum_{r=0}^RN^{+}_{j,r}}{\sum_{r=0}^RN_{j,r}},\quad\text{recall}@R=\frac{1}{M}\sum_{j=1}^M\frac{\sum_{r=0}^RN^{+}_{j,r}}{N^{+}_{j}},\nonumber
\end{align}

\section{Radius Aware Mean Average Precision}
In this section, we first analyze the problems of existing metrics. Then, we propose a new metric for evaluating hash codes.

\subsection{Observations}\label{sec:obs}
\begin{figure*}[tb]
\centering
\begin{minipage}[t]{0.48\linewidth}
\centering
\begin{tabular}{c@{ }@{ }c@{ }@{ }c@{ }@{ }c}
    \begin{minipage}[t]{0.49\linewidth}
    \centering
    \includegraphics[width=1\textwidth]{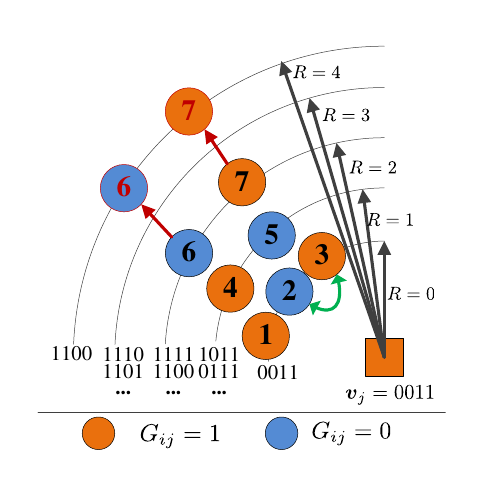} \\
    (a) code \#1
    \end{minipage}
    \begin{minipage}[t]{0.49\linewidth}
    \centering
    \includegraphics[width=1\textwidth]{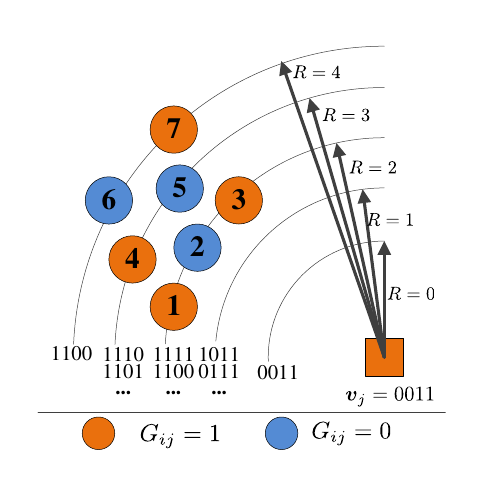} \\
    (b) code \#2
    \end{minipage}
\end{tabular}
\caption{Problem of MAP.}
\label{fig:iMAP}
\end{minipage}
\hspace{0.1cm}
\begin{minipage}[t]{0.48\linewidth}
\centering
\begin{tabular}{c@{ }@{ }c@{ }@{ }c@{ }@{ }c}
    \begin{minipage}[t]{0.49\linewidth}
    \centering
    \includegraphics[width=1\textwidth]{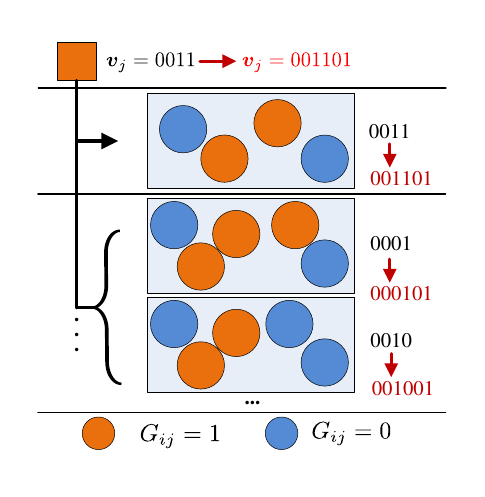} \\
    (a) code \#1
    \end{minipage}
    \begin{minipage}[t]{0.49\linewidth}
    \centering
    \includegraphics[width=1\textwidth]{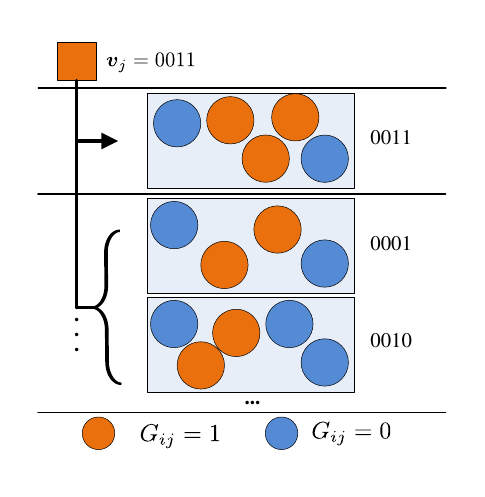} \\
    (b) code \#2
    \end{minipage}
\end{tabular}
\caption{Problem of precision and recall.}
\label{fig:iPreRec}
\end{minipage}
\end{figure*}

We find that there are some problems when we utilize MAP to evaluate a hashing algorithm. We present an example in Figure~\ref{fig:iMAP}. In Figure~\ref{fig:iMAP}, we use an orange square to denote the query, i.e., $\v_j$. The semicircles denote the different Hamming distance away from the query. We design two coding strategies, i.e., ``code \#1'' and ``code \#2'', which are shown as data points from data \#1 to data \#7 in Figure~\ref{fig:iMAP}~(a) and Figure~\ref{fig:iMAP}~(b), respectively. Different colors are used to denote if a data point is a similar data~(orange) of the query or not~(blue). The number in each data point denotes the position of the data point in the ranked list. 

The first problem is that MAP suffers from uncertainty problem~\cite{DBLP:conf/cvpr/0003CBS18} as the ranked list is based on \emph{integer-valued} Hamming distance. Notice that the data points in the same $r$-Hamming ball are essentially unordered. If we change the position of these data points, e.g., example shown in green arrow in Figure~\ref{fig:iMAP}~(a), we can get different MAP values. The second problem is that MAP only depends on \emph{relative} Hamming distance. For example, the MAP values for code \#1 and code \#2 are exactly the same. However, code \#1 is obviously better than code \#2. The last problem is that the MAP ignores the retrieval time cost. For example, if we move the data \#6 and data \#7 to the fifth semicircle for code \#1~(shown in red arrow in Figure~\ref{fig:iMAP}~(a)), the MAP value will not change. However, we must spend more time to retrieve them when the Hamming distance away from the query is enlarged.

Although precision and recall can avoid uncertainty problem and don't depend on relative Hamming distance, there still exist some problems for these metrics. For precision and recall, we also present an example to show the weaknesses of them in Figure~\ref{fig:iPreRec}. In Figure~\ref{fig:iPreRec}, the notations of query and coding strategies are defined similarly to the notations defined in Figure~\ref{fig:iMAP}. 

The first problem is that when we calculate the precision@$R$~(or recall@$R$), the data points within \emph{different} $r$-Hamming balls, where $0\le r<R$, are totally unordered. We point out that to avoid uncertainty problem, the data points within the \emph{same} $r$-Hamming ball should be unordered, but the data points in \emph{different} $r$-Hamming balls should be ordered. In other words, precision@$R$ and recall@$R$ fail to evaluate global accuracy. For example, for \mbox{code \#1}, $N_{j,0}^{+}=2,N_{j,1}^{+}=5$, and for \mbox{code \#2}, $N_{j,0}^{+}=3,N_{j,1}^{+}=4$. However, we can find that $\text{precision@}1=0.5$ and $\text{recall}@1=0.7$~(assume that $N^{+}_j=10$) for both strategies. Hence, for a specific Hamming radius $R$, we might can't differentiate which hash codes is a superior one based on precision@$R$ or recall@$R$. The second problem is that precision and recall ignore the retrieval time cost. For example, if we add redundant and exactly same $s$-bits to the query $\v_j$ and the database hash codes for query in \mbox{code \#1}~(shown as red hash codes in Figure~\ref{fig:iPreRec}~(a)), the precision and recall at radius $R$ will not change. However, the retrieval time cost will increase obviously.

\subsection{Radius Aware Mean Average Precision}
\begin{table*}[tb]
\centering
\caption{Statistic for hash lookup based retrieval.}
\label{tab:stat}
 \begin{tabular}{l|c|c|c|c}\hline
 Bins & \#data points & \#ground-truth  & Time cost & Precision@$R$\\\hline
$\{\b^{(0)}_{j_1}\}$	
                & $N_{j,0}$ & $N_{j,0}^+$ & $t_0$ & $\frac{N_{j,0}^+}{N_{j,0}}$\\\hline
$\{\b^{(1)}_{j_1},\b^{(1)}_{j_2},\dots,\b^{(1)}_{j_{C_Q^1}}\}$
                & $N_{j,1}$ & $N_{j,1}^+$ & $(1+C_Q^1)t_0$ & $\frac{N_{j,0}^++N_{j,1}^+}{N_{j,0}+N_{j,1}}$\\\hline
$\cdots$       & $\cdots$ & $\cdots$ & $\cdots$ & $\cdots$\\\hline
$\{\b^{(R)}_{j_1},\b^{(R)}_{j_2},\dots,\b^{(R)}_{j_{C_Q^R}}\}$
                & $N_{j,R}$ & $N_{j,R}^+$ & $(\sum_{r=0}^R C_Q^r)t_0$ & $\frac{\sum_{r=0}^RN_{j,r}^+}{\sum_{r=0}^RN_{j,r}}$\\\hline
                $\cdots$       & $\cdots$ & $\cdots$ & $\cdots$ & $\cdots$\\\hline
$\{\b^{(Q)}_{j_1}\}$
                & $N_{j,Q}$ & $N_{j,Q}^+$ & $(\sum_{r=0}^Q C_Q^r)t_0$ & $\frac{\sum_{r=0}^QN_{j,r}^+}{\sum_{r=0}^QN_{j,r}}$\\\hline
\end{tabular}
\end{table*}
According to the aforementioned observations, a proper evaluation metric for hashing should be able to avoid uncertainty problem and evaluate global accuracy. Furthermore, as the retrieval time cost is an important factor reflecting the performance of search, it should be considered.

We first present some related information about the precision at Hamming radius $R$ for a given query $\v_j$ in Table~\ref{tab:stat}. Then we have:
\begin{align}
\text{precision}(\y_j,R)=\frac{\sum_{r=0}^RN_{j,r}^+}{\sum_{r=0}^RN_{j,r}}.\nonumber
\end{align}

To introduce the effect of the retrieval time cost, we impose the time cost penalty on all precision@$R$. Specifically, when we retrieve to $R$-Hamming ball, the time cost is $(\sum_{r=0}^R C_Q^r)t_0$, we multiply precision$(\y_j,R)$ by the rate of $t_0$ and $(\sum_{r=0}^R C_Q^r)t_0$. Then we reformulate the precision as follows:
\begin{align}
\text{precision}(\y_j,R)=\frac{\sum_{r=0}^RN_{j,r}^+}{\sum_{r=0}^RN_{j,r}}\cdot\frac{t_0}{(\sum_{r=0}^R C_Q^r)t_0}
=\frac{\sum_{r=0}^RN_{j,r}^+}{\sum_{r=0}^RN_{j,r}}\cdot\frac{1}{\sum_{r=0}^R C_Q^r}.\nonumber
\end{align}

Then we define the radius aware mean average precision as follows:
\begin{align}
\text{RAAP}(\y_j,R)=\frac{\sum_{r=0}^R\text{precision}(\y_j,r)}{R+1},\quad
\text{RAMAP}@R=\frac{1}{M}\sum_{j=1}^M\text{RAAP}(\y_j,R).\label{eq:ramap}
\end{align}
Based on the definition of RAMAP in~(\ref{eq:ramap}), we can find that our proposed metric has the following advantages:
\begin{inparaenum}[a)]
\item RAMAP considers the effect of retrieval time cost and imposes the time cost penalty on the metric.
\item RAMAP can avoid uncertainty problem as it only depends on accuracy at Hamming radius.
\item RAMAP doesn't depend on relative Hamming distance. Hence, RAMAP is more proper for evaluating hash codes.
\item RAMAP averages the precision at all Hamming radius $r\le R$ when we calculate RAMAP@$R$. Thus RAMAP can provide more proper evaluation for hash codes.
\end{inparaenum}
\section{Coding Strategy for Metric Comparison}
To verify the superiority of RAMAP, we propose two coding strategies in this section, including a heuristic coding strategy and a learning based coding strategy. Through these two coding strategies, we point out that existing metrics can't provide a proper evaluation for hashing algorithm.
\subsection{Heuristic Coding Strategy}\label{sec:HE}
\paragraph{Methodology}
Based on the observations in Section~\ref{sec:obs}, we heuristically design a coding strategy to compare RAMAP with existing metrics. Specifically, given $M$ query binary codes $\V=\{\v_j\}_{j=1}^M$ and $N$ database binary codes $\U=\{\u_i\}_{i=1}^N$. We define two transformation methods to convert the $\U$ and $\V$ to extension codes as follows:
\begin{align}
&\forall i\in\{1,\dots,n\},\bar\u_{i}=[\u_{i};\bone],\quad\forall j\in\{1,\dots,m\},\bar\v_{j}=[\v_{j};\bone],\label{hc:se}\\
&\forall i\in\{1,\dots,n\},\widehat\u_i=[\u_i;\bzero],\quad\forall j\in\{1,\dots,m\},\widehat\v_j=[\v_j;\bone],\label{hc:de}
\end{align}
where $\bone\in\{1\}^s,\bzero\in\{0\}^s$ and $[\a;\b]$ denotes the vector concatenation operation. Then we can get three hash codes sets, i.e., $\{\U,\V\}$, $\{\bar\U,\bar\V\}$ and $\{\widehat\U,\widehat\V\}$. We name the first transformation method in~(\ref{hc:se}) as \emph{same-extension method} and the second method in~(\ref{hc:de}) as \emph{different-extension method}.

According to the definition of these hash codes, given any hash codes pairs $\{\u_i,\v_j\}$, $\{\bar\u_i,\bar\v_j\}$ and $\{\widehat\u_i,\widehat\v_j\}$, we have the following equation:
\begin{align}
\text{dist}_H(\u_i,\v_j)=\text{dist}_H(\bar\u_i,\bar\v_j)=\text{dist}_H(\widehat\u_i,\widehat\v_j)-s.\nonumber
\end{align}

\paragraph{Analysis} According to the definition of these binary codes, we have the following observations:
\begin{inparaenum}[a)]
\item The coding strategy $\{\U,\V\}$ is better than $\{\bar\U,\bar\V\}$ as the latter contains redundant $s$-bits without any information.
\item The coding strategy $\{\bar\U,\bar\V\}$ is better than  $\{\widehat\U,\widehat\V\}$ as the Hamming distance in the latter hash codes is larger than the former.
\item The MAP values for these three coding strategies will be exactly the same.
\item The precision and recall for coding strategy $\{\U,\V\}$ and $\{\bar\U,\bar\V\}$ will be exactly the same.
\item If we didn't impose the time cost penalty on precision@$R$ when we design the RAMAP, the RAMAP for $\{\U,\V\}$ and $\{\bar\U,\bar\V\}$ will be exactly the same.
\item According to the definition of RAMAP, we can find that the RAMAP of coding strategy $\{\U,\V\}$ is better than that of $\{\bar\U,\bar\V\}$, and the RAMAP of coding strategy $\{\bar\U,\bar\V\}$ is better than that of $\{\widehat\U,\widehat\V\}$.
\end{inparaenum}
\subsection{Learning based Coding Strategy}\label{sec:LE}
To further demonstrate the superiority of our proposed RAMAP, we propose a learning based coding strategy to compare RAMAP with MAP, precision. Please note that in this section, the hash codes are redefined on $\{-1,+1\}^Q$ for simplicity, i.e., $\H\doteq 2\H-1$, after learning procedure, all hash codes can be converted to the form of $\{0,1\}^Q$.
\paragraph{Model}
Given a dataset $\Z=\{\z_i\}_{k=1}^L$ and the similarity $\S=\{S_{kl}\}_{k,l=1}^L$, we adopt widely used squared loss~\cite{DBLP:conf/icml/WangKC10,DBLP:conf/cvpr/LiuWJJC12} to learn similarity preserving hash codes. The squared loss is defined as follows:
\begin{align}
\min_{\H}\;\LM(\H;\Z,\S)&=\frac{1}{L^2}\sum_{k,l=1,k\neq l}^L\big[(Q-QS_{kl})-\text{dist}_H(\h_k,\h_l)\big]^2\label{eq:twoL}\\
&=\underbrace{\frac{1}{\vert\S^+\vert}\sum_{S_{kl}=1,k\neq l}\big[\text{dist}_H(\h_k,\h_l)\big]^2}_{\LM^{+}(\H)}+\underbrace{\frac{1}{\vert\S^-\vert}\sum_{S_{kl}=0,k\neq l}\big[Q-\text{dist}_H(\h_k,\h_l)\big]^2}_{\LM^{-}(\H)},\nonumber\\
\text{subject to:}\;&\forall k,\h_k\in\{-1,+1\}^Q,\h_k=f(\z_k),\nonumber
\end{align}
where $\S^+=\{S_{kl}\;\vert\; \forall S_{kl}=1, S_{kl}\in\S\},\S^-=\{S_{kl}\;\vert\;\forall S_{kl}=0, S_{kl}\in\S\}$.

We utilize a deep neural network~(DNN) to map the data points into $\RB^Q$. Specifically, we adopt pre-trained Alexnet~\cite{DBLP:conf/nips/KrizhevskySH12} on ImageNet~\cite{DBLP:conf/nips/KrizhevskySH12} dataset as a backbone network and replace the last fully-connection layer as a hash layer which projects the 4,096 features into $\RB^Q$. We use $g(\z;\Theta)$ to denote the output of DNN for input $\z$, where $\Theta$ denotes the parameters of DNN. Then we define the hash function as $f(\z;\Theta)=\sgn(g(\z;\Theta))$. We name this method as \emph{deep squared hashing}~(DSH).

From problem~(\ref{eq:twoL}), we can see that the core idea of squared loss is to pull similar pair closer and push the dissimilar pair farther through minimizing $\LM^+(\H)$ and $\LM^{-}(\H)$, respectively. That is to say, if we change the $\LM^{+}(\H)$ as : $\JM^{+}(\H;m)=\frac{1}{\vert\S^+\vert}\sum_{S_{kl}=1,k\neq l}\big[m-\text{dist}_H(\h_k,\h_l)\big]^2$, where $m\ll Q$. The relative Hamming distance will not change by minimizing $\JM(\H;\Z,\S,m)=\JM^{+}(\H;m)+\LM^{-}(\H)$. Hence, we construct a new approach by solving the following problem:
\begin{align}
\min_{\H}\;\JM(\H;\Z,\S,m)&=\JM^{+}(\H;m)+\LM^{-}(\H);\quad
\text{subject to:}\;\forall k,\h_k\in\{-1,+1\}^Q,\h_k=f(\z_k).\nonumber
\end{align}

Like DSH, we adopt a modified pre-trained Alexnet to map input data into $\RB^Q$. Then a sign function is adopted to generate hash codes. We call this method as \emph{$m$-deep squared hashing}~($m$-DSH).

\paragraph{Learning} As the existing of sign function, we can't adopt back-propagation~(BP) algorithm to learn the parameters of DNN. Hence we replace $\sgn(x)$ as $\tanh(\beta x)$~\cite{DBLP:conf/iccv/CaoLWY17}. Then we can re-formulate the objective function in~(\ref{eq:twoL}) as follows:
\begin{align}
\min_{\widetilde\H}\;\LM(\widetilde\H;\Z,\S)&=\LM^{+}(\widetilde\H)+\LM^{-}(\widetilde\H);\quad\text{subject to:}\;\forall k,\widetilde\h_k=\tanh(\beta g(\z_k;\Theta)).\nonumber
\end{align}
Then we adopt a mini-batch based BP algorithm to learn parameters $\Theta$. The learning algorithm for $m$-DSH can be derived similarly. After learning, we utilize the equation $\v=\sgn(g(\y;\Theta))$ to generate binary code for unseen sample $\y$.

\paragraph{Analysis} Compared DSH with $m$-DSH, we can find that:
\begin{inparaenum}[a)]
\item The goal of DSH is trying to map the Hamming distance $\text{dist}_H(\h_k,\h_l)$ into $[0,Q]$ based on $S_{kl}$. If $S_{kl}=1$, the $\text{dist}_H(\h_k,\h_l)$ is pulled to 0.
\item The goal of $m$-DSH is trying to map the Hamming distance $\text{dist}_H(\h_k,\h_l)$ into $[m,Q]$ based on $S_{kl}$. If $S_{kl}=1$, the $\text{dist}_H(\h_k,\h_l)$ is pulled to $m$.
\item Both of DSH and $m$-DSH are relative Hamming distance preserving hashing method.
\item We can find that the MAP of DSH and $m$-DSH will be very close.
\item We can find that the RAMAP of DSH will be better than that of $m$-DSH.
\item As $m$-DSH can preserve relative Hamming distance, the precision@$R=m$ of $m$-DSH might be as high as precision@$R=0$ of DSH.
\end{inparaenum}

\section{Experiments Analysis}
To verify the superiority of our proposed metric, we carry out experiments based on two proposed coding strategies. The experiments are run on a workstation with Intel (R) Xeon(R) CPU E5-2620V4@2.1G of 8 cores, 128G RAM and an NVIDIA TITAN XP GPU.
\subsection{Experimental Settings}
We adopt CIFAR10~\cite{krizhevsky2009learning} for our experiments. CIFAR10 dataset contains 60,000 32$\times$32 images which belong to 10 classes. Following the setting of existing hashing algorithms~\cite{DBLP:conf/nips/LiSHT17,DBLP:conf/nips/SuZHT18}, we utilize 1,000 images~(100 images per class) as query set and the remaining images as database set. For this dataset, two images will be defined as a ground-truth neighbor~(similar pair) if they share the common label.

For the heuristic coding strategy, we utilize 4,096-dimensional deep features which are extracted by the pre-trained Alexnet~\cite{DBLP:conf/nips/KrizhevskySH12} model on ImageNet~\cite{DBLP:conf/nips/KrizhevskySH12} dataset . We adopt locality sensitive hashing~(LSH)~\cite{DBLP:conf/compgeom/DatarIIM04} to obtain basic hash codes. We set $Q=8$ and $s=1,8$. And then we obtain the $\{\bar\U,\bar\V\}$ and $\{\widehat\U,\widehat\V\}$ based on $\{\U,\V\}$.

For learning based coding strategy, 5,000~(500 images per class) images are randomly sampled from database set to construct training set. And we resize all images to $224\times 224$ and use the raw pixels as the inputs. We utilize a pre-trained Alexnet as the backbone network. We set initial learning rate as 0.05 and reduce it to 0.025 after 200 epochs~(we train our algorithms 400 epochs totally). We set weight-decay as $5\times 10^{-5}$ and mini-batch size as 128. $\beta$ will be enlarged by a factor of 1.005 per epoch to reduce quantization error. For these methods, all experiments are run five times with different random seeds and average accuracy is reported.
\subsection{Experiments Results}
\paragraph{Results for Heuristic Coding Strategy} For the heuristic coding strategy, we compare RAMAP with MAP and precision, recall in Table~\ref{tab:ramap-map} and Table~\ref{tab:ramap-pre-rec} respectively. In Table~\ref{tab:ramap-map} and Table~\ref{tab:ramap-pre-rec}, we utilize ``LSH'' to denote the hash codes generated by LSH algorithm. We utilize ``LSH$_{SE}$'' and ``LSH$_{DE}$'' to denote the same-extension method and different-extension method, respectively. We omit some experiments results, e.g., RAMAP@$R=5$, in these tables due to space limitation.

In Table~\ref{tab:ramap-map}, we compare the RAMAP with MAP. We can see that the MAP values for LSH, LSH$_{SE}$ and LSH$_{DE}$ are exactly the same. That is to say, we can't decide which method is better according to MAP values. However, from RAMAP values, we can see that LSH and LSH$_{SE}$ are better than LSH$_{DE}$. We can also find that LSH is better than LSH$_{SE}$ slightly.
\begin{table*}[tb]
\centering
\caption{RAMAP and MAP for heuristic coding strategy on CIFAR10 dataset.}
\label{tab:ramap-map}
\begin{tabular}{l|c|c|c|c|c|c|c|c|c}
\Xcline{1-10}{0.8pt}
\multirow{2}{*}{Metric}  &\multirow{2}{*}{Method} & \multicolumn{8}{c}{\#returned samples/Radius}\\\cline{3-10}
&            & 5K/0   & 10K/1  & 15K/2  & 20K/3  & 25K/4  & $\dots$ & 45K/8  & 50K/9 \\\hline
\multirow{3}{*}{MAP}
&LSH         & 0.1792 & 0.1644 & 0.1624 & 0.1521 & 0.1501 & $\dots$ & 0.1375 & 0.1365\\\cline{2-10}
&LSH$_{SE}$  & 0.1792 & 0.1644 & 0.1624 & 0.1521 & 0.1501 & $\dots$ & 0.1375 & 0.1365\\\cline{2-10}
&LSH$_{DE}$  & 0.1792 & 0.1644 & 0.1624 & 0.1521 & 0.1501 & $\dots$ & 0.1375 & 0.1365\\\hline
\multirow{3}{*}{RAMAP}
&LSH         & 0.1896 & 0.1040 & 0.0706 & 0.0533 & 0.0428 & $\dots$ & 0.0239 & N/A   \\\cline{2-10}
&LSH$_{SE}$  & 0.1896 & 0.1030 & 0.0697 & 0.0525 & 0.0421 & $\dots$ & 0.0235 & 0.0212\\\cline{2-10}
&LSH$_{DE}$  & 0.0000 & 0.0095 & 0.0075 & 0.0059 & 0.0048 & $\dots$ & 0.0028 & 0.0025\\
\Xcline{1-10}{0.8pt}
\end{tabular}
\end{table*}

In Table~\ref{tab:ramap-pre-rec}, we compare RAMAP with precision and recall for LSH, LSH$_{SE}$. We can find that precision, recall for LSH and LSH$_{SE}$ are exactly the same. However, based on the analysis in Section~\ref{sec:HE}, LSH is better than LSH$_{SE}$. In other words, based on precision and recall, we can't decide which hash codes is better in this case. Based on RAMAP, we can find that LSH is better than LSH$_{SE}$, which conforms to the analysis in Section~\ref{sec:HE}.
\begin{table*}[tb]
\centering
\caption{RAMAP, precision and recall for heuristic coding strategy on CIFAR10 dataset.}
\label{tab:ramap-pre-rec}
\begin{tabular}{l|c|c|c|c|c|c|c|c|c}
\Xcline{1-10}{0.8pt}
\multirow{2}{*}{Metric}  &\multirow{2}{*}{Method} &\multirow{2}{*}{$s$} & \multicolumn{7}{c}{Radius}\\\cline{4-10}
&            &        & $R=0$  & $R=1$  & $R=2$  & $R=3$  & $\dots$ & $R=8$  & $R=9$ \\\hline
\multirow{3}{*}{Precision}
&LSH         & N/A    & 0.1896 & 0.1651 & 0.1433 & 0.1261 & $\dots$ & 0.1000 & N/A   \\\cline{2-10}
&LSH$_{SE}$  & $s=1$  & 0.1896 & 0.1651 & 0.1433 & 0.1261 & $\dots$ & 0.1000 & 0.1000\\\cline{2-10}
&LSH$_{SE}$  & $s=8$  & 0.1896 & 0.1651 & 0.1433 & 0.1261 & $\dots$ & 0.1000 & 0.1000\\\hline
\multirow{3}{*}{Recall}
&LSH         & N/A    & 0.0090 & 0.0637 & 0.2147 & 0.4609 & $\dots$ & 1.0000 & N/A   \\\cline{2-10}
&LSH$_{SE}$  & $s=1$  & 0.0090 & 0.0637 & 0.2147 & 0.4609 & $\dots$ & 1.0000 & 1.0000\\\cline{2-10}
&LSH$_{SE}$  & $s=8$  & 0.0090 & 0.0637 & 0.2147 & 0.4609 & $\dots$ & 1.0000 & 1.0000\\\hline
\multirow{3}{*}{RAMAP}
&LSH         & N/A    & 0.1896 & 0.1040 & 0.0706 & 0.0533 & $\dots$ & 0.0239 & N/A   \\\cline{2-10}
&LSH$_{SE}$  & $s=1$  & 0.1896 & 0.1030 & 0.0697 & 0.0525 & $\dots$ & 0.0235 & 0.0212\\\cline{2-10}
&LSH$_{SE}$  & $s=8$  & 0.1896 & 0.0996 & 0.0668 & 0.0501 & $\dots$ & 0.0223 & 0.0201\\
\Xcline{1-10}{0.8pt}
\end{tabular}
\end{table*}

\begin{figure*}[tb]
\centering
\begin{minipage}[t]{0.48\linewidth}
\centering
\begin{tabular}{c@{ }@{ }c@{ }@{ }c@{ }@{ }c}
    \begin{minipage}[t]{0.49\linewidth}
    \centering
    \includegraphics[width=1\textwidth]{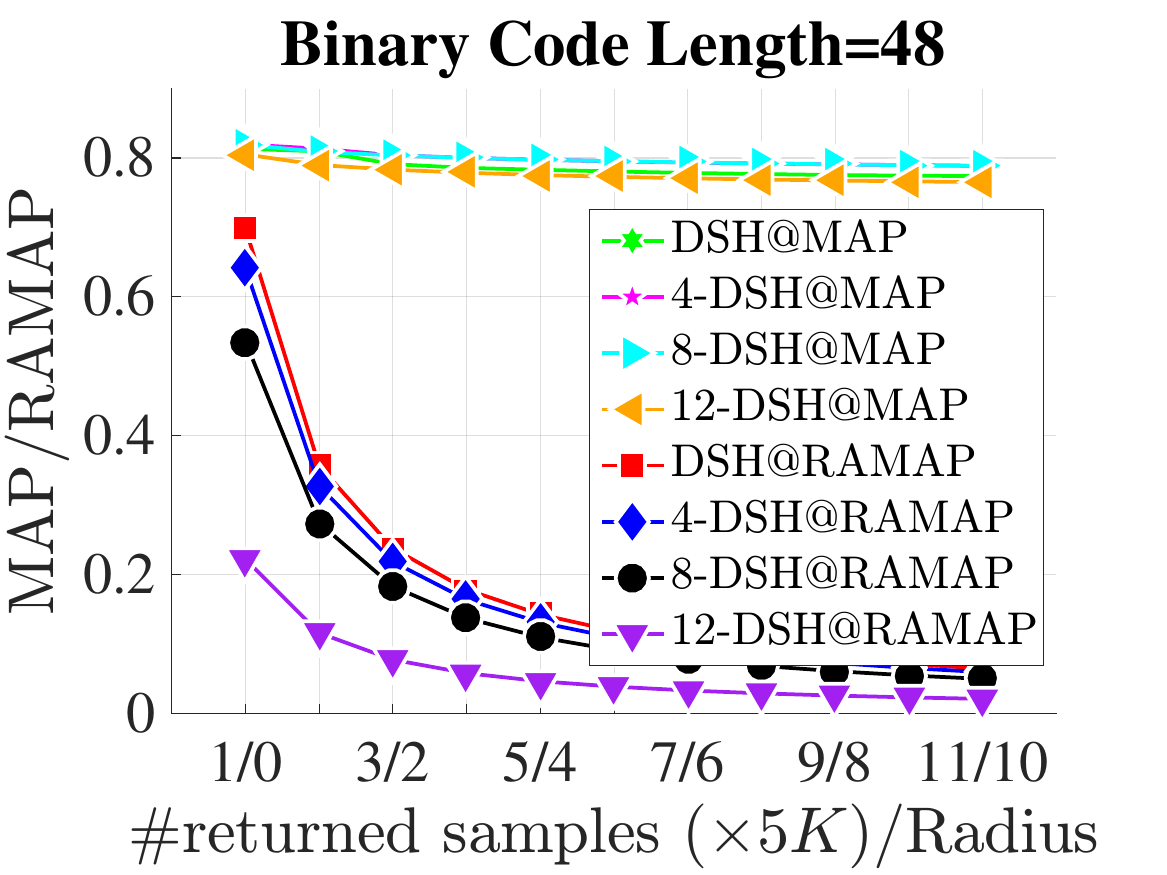} \\
    (a). 48 bits
    \end{minipage}
    \begin{minipage}[t]{0.49\linewidth}
    \centering
    \includegraphics[width=1\textwidth]{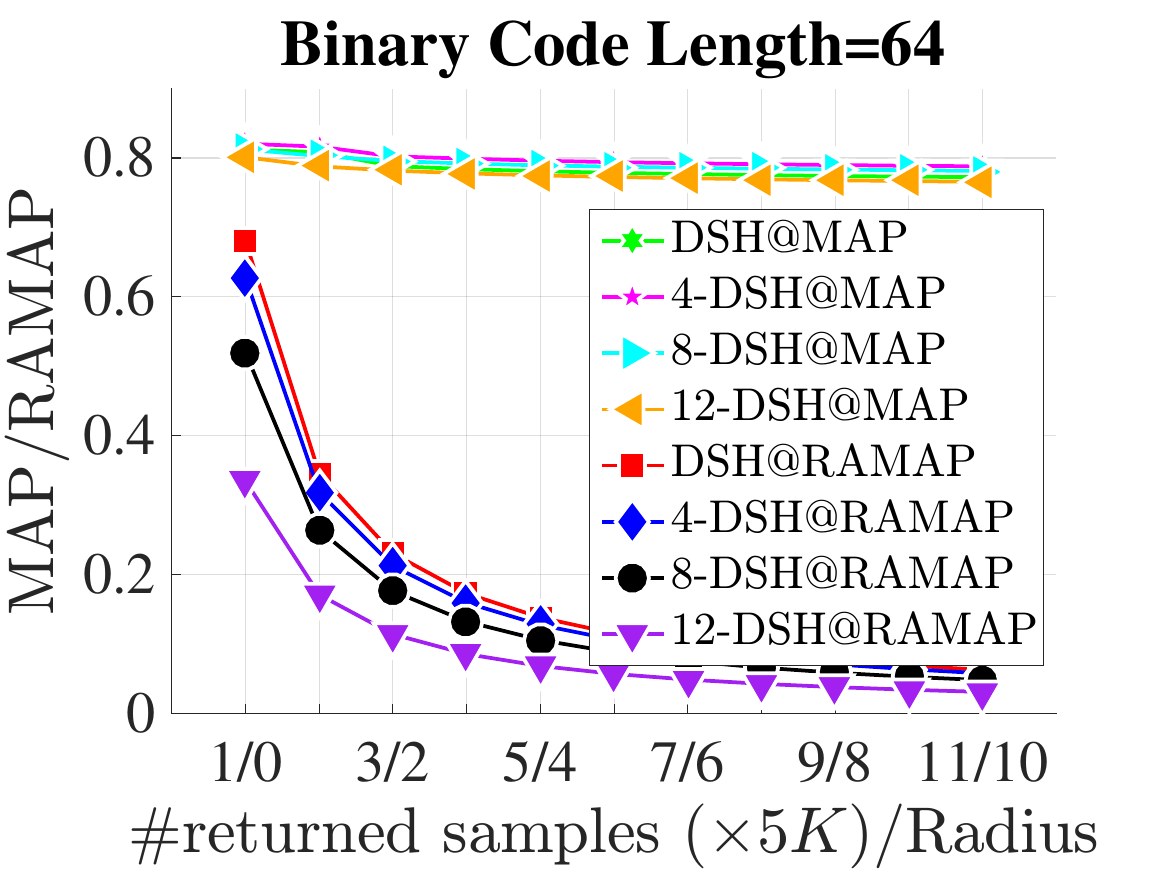} \\
    (b). 64 bits
    \end{minipage}
\end{tabular}
\caption{RAMAP vs. MAP.}
\label{fig:ramap-map}
\end{minipage}
\hspace{0.1cm}
\begin{minipage}[t]{0.48\linewidth}
\centering
\begin{tabular}{c@{ }@{ }c@{ }@{ }c@{ }@{ }c}
    \begin{minipage}[t]{0.49\linewidth}
    \centering
    \includegraphics[width=1\textwidth]{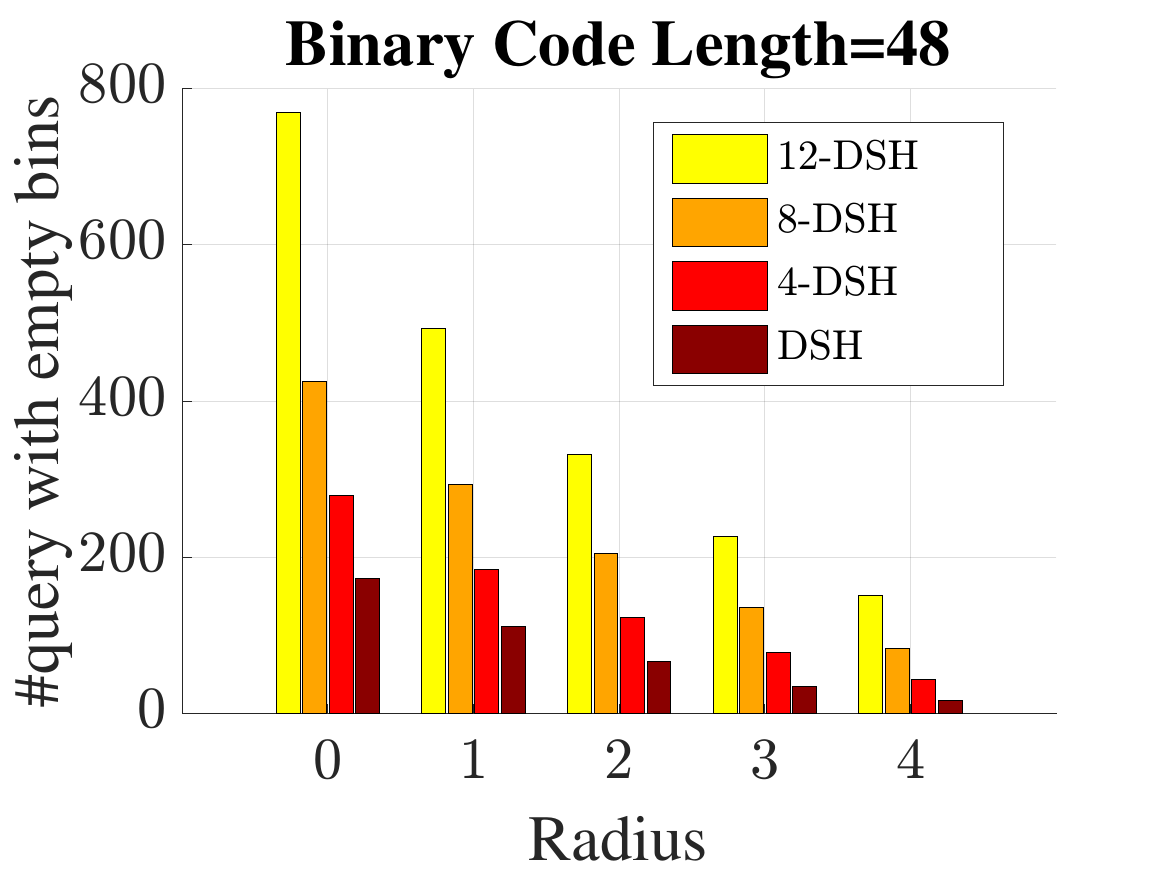} \\
    (a). 48 bits
    \end{minipage}
    \begin{minipage}[t]{0.49\linewidth}
    \centering
    \includegraphics[width=1\textwidth]{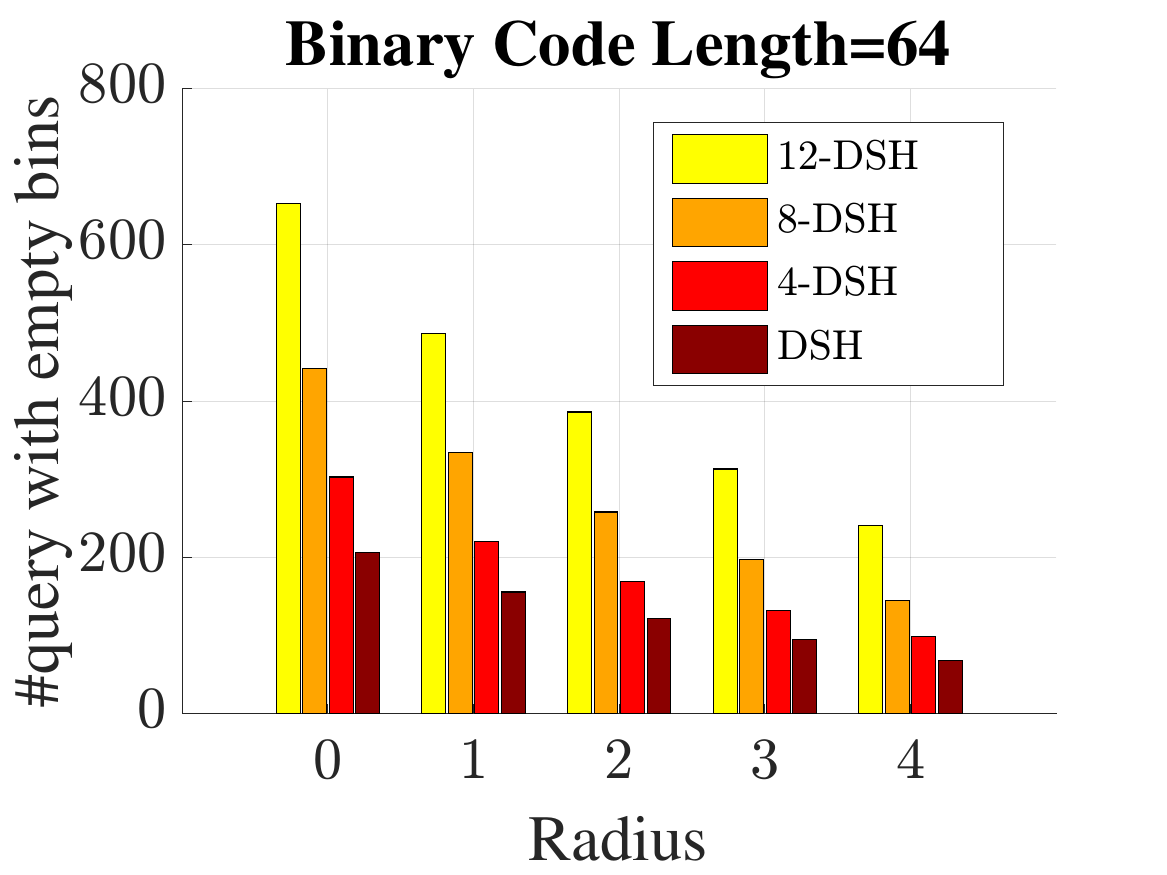} \\
    (b). 64 bits
    \end{minipage}
\end{tabular}
\caption{\#query with empty bins.}
\label{fig:num-empty-bins}
\end{minipage}
\end{figure*}

\begin{figure*}[tb]
\centering
\begin{minipage}[t]{0.48\linewidth}
\centering
\begin{tabular}{c@{ }@{ }c@{ }@{ }c@{ }@{ }c}
    \begin{minipage}[t]{0.49\linewidth}
    \centering
    \includegraphics[width=1\textwidth]{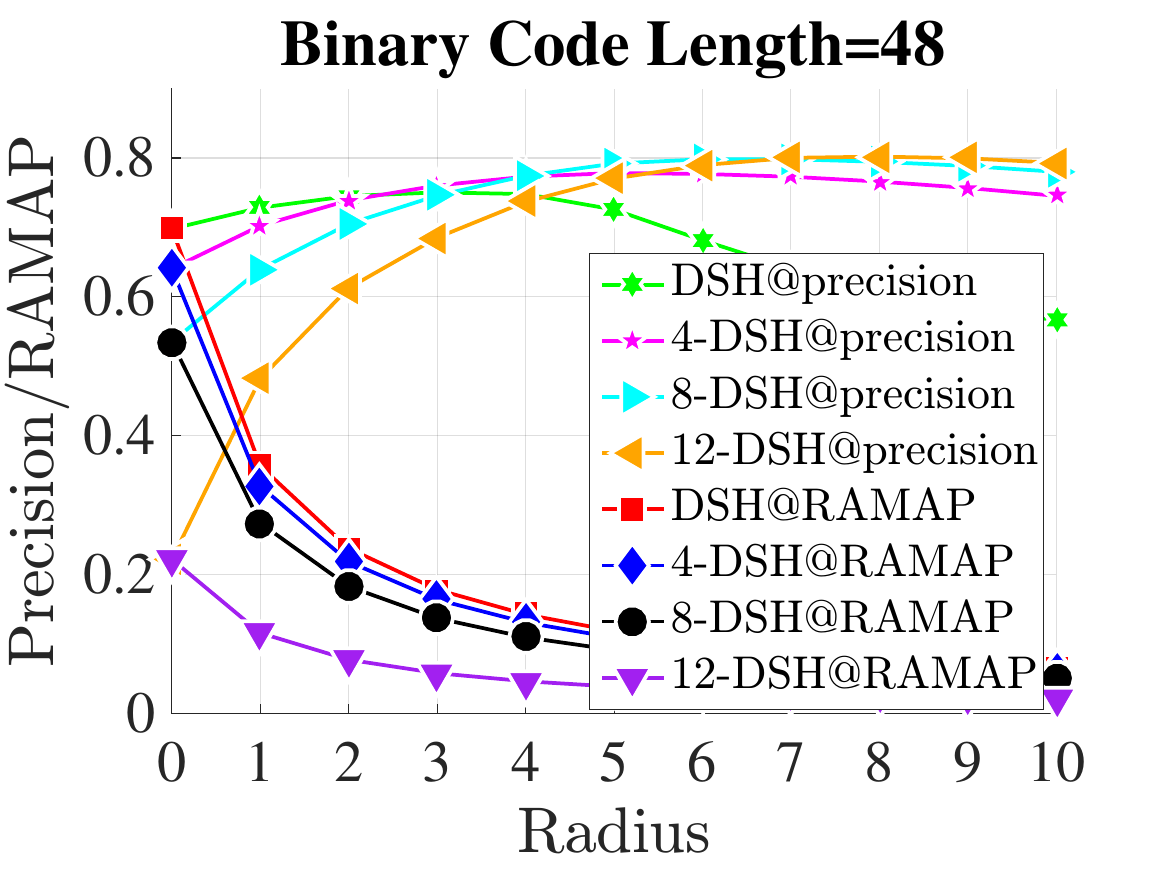} \\
    (a). 48 bits
    \end{minipage}
    \begin{minipage}[t]{0.49\linewidth}
    \centering
    \includegraphics[width=1\textwidth]{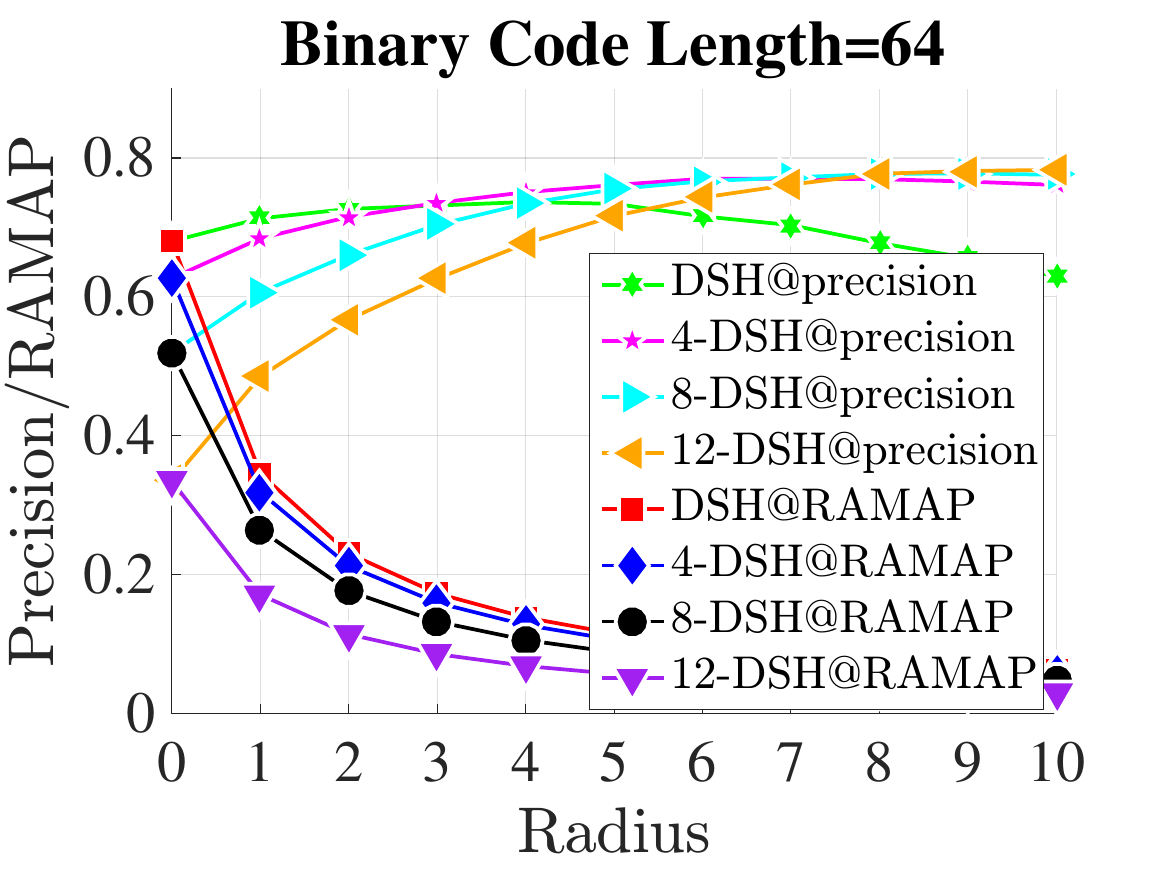} \\
    (b). 64 bits
    \end{minipage}
\end{tabular}
\caption{RAMAP vs. precision.}
\label{fig:ramap-pre}
\end{minipage}
\hspace{0.1cm}
\begin{minipage}[t]{0.48\linewidth}
\centering
\begin{tabular}{c@{ }@{ }c@{ }@{ }c@{ }@{ }c}
    \begin{minipage}[t]{0.49\linewidth}
    \centering
    \includegraphics[width=1\textwidth]{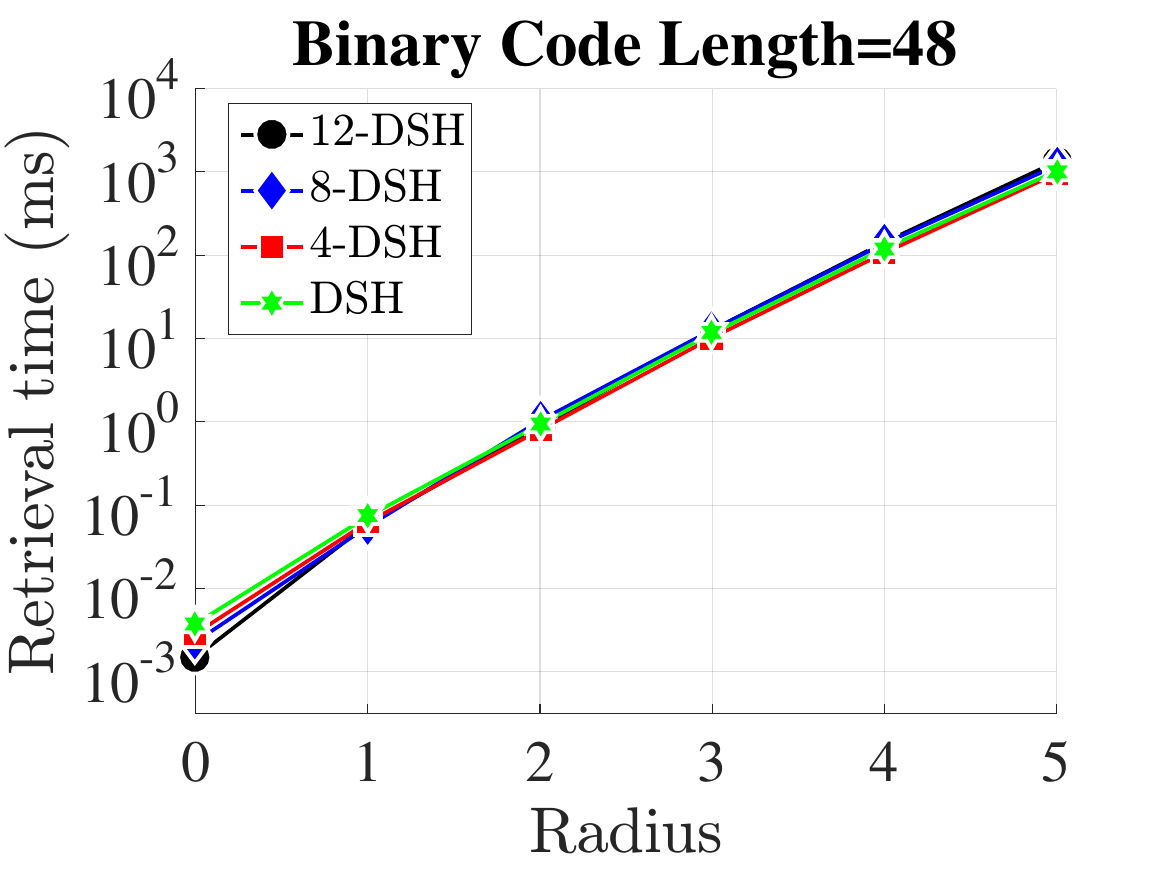} \\
    (a). 48 bits
    \end{minipage}
    \begin{minipage}[t]{0.49\linewidth}
    \centering
    \includegraphics[width=1\textwidth]{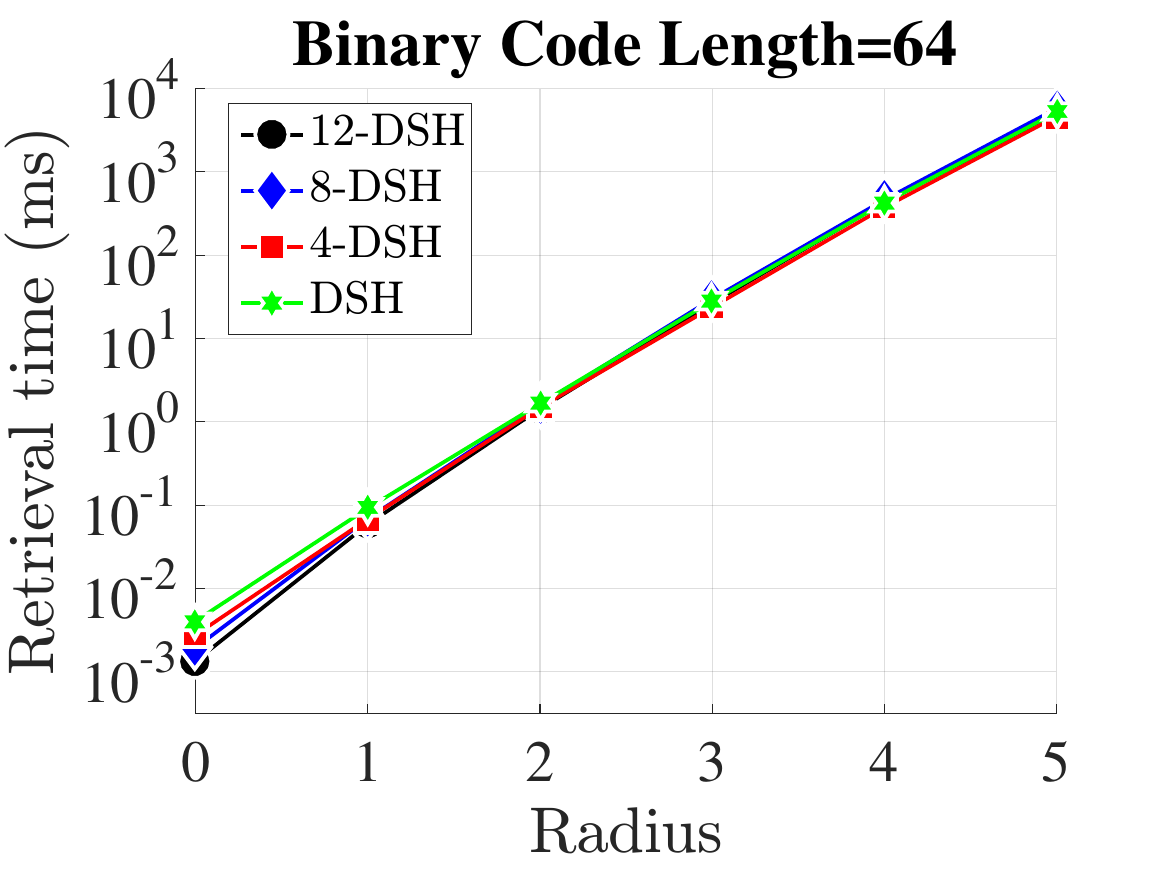} \\
    (b). 64 bits
    \end{minipage}
\end{tabular}
\caption{Retrieval time.}
\label{fig:time}
\end{minipage}
\end{figure*}

\paragraph{Results for Learning based Coding Strategy} For the learning based coding strategy, we set $m=4,8,12$ for this experiment and we use ``$4$-DSH'', ``$8$-DSH'' and ``$12$-DSH'' to denote the corresponding $m$-DSH method.

In Figure~\ref{fig:ramap-map}, we present the RAMAP and MAP for learning based coding strategy with binary code length being 48 bits and 64 bits. From Figure~\ref{fig:ramap-map}, we can find that the MAP for all methods are very close. Hence, we can't decide which one is the best algorithm based on MAP. However, according to the RAMAP, we can clearly find that the best algorithm is DSH, the second best is $4$-DSH, the third best is $8$-DSH and the worst is $12$-DSH. In Figure~\ref{fig:num-empty-bins}, we present the number of query with empty bins for these methods with different Hamming radius. We can find that the retrieval procedure will suffer from the worst empty-bin problem if we adopt the learned hash codes of $12$-DSH to construct inverted index table. That is to say, the worst method is $12$-DSH, the second worst method is $8$-DSH, the third worst method is $4$-DSH and the best method is DSH, which is in accordance with the results of RAMAP and the analysis in Section~\ref{sec:LE}.

Furthermore, we compare RAMAP with precision at Hamming radius $R$ with binary code length being 48 bits and 64 bits in Figure~\ref{fig:ramap-pre}. From Figure~\ref{fig:ramap-pre}, we can see that although the precision values with low radiuses are differentiable, we still might be confused to choose a better algorithm as the precision values with high radiuses lead to confusion. For example, although the precision@$R=0,1,2$ values of $4$-DSH are lower than that of DSH, the precision values of $4$-DSH are higher than DSH when $R\ge 3$. Thus we will be confused to decide which one is better according to precision. While based on the RAMAP of these methods, we can find that DSH is better than $4$-DSH, $4$-DSH is better than $8$-DSH and $12$-DSH is the worst method. The reason why precision can't provide deterministic conclusion might be two-fold. On one hand, based on the model of $m$-DSH, the precision@$R=m$ might be as high as the precision@$R=0$ of DSH. However, precision@$R=m$ ignores the precision@$R=\{0,\dots,m-1\}$, i.e., precision fails to evaluate global performance. On the other hand, the precision ignores retrieval time. To verify this point, we utilize the learned hash codes to construct inverted index table and perform bucket search. We report the time cost~(in millisecond) in Figure~\ref{fig:time}. From Figure~\ref{fig:time}, we can see that with the increasing of search radius, the time cost increases exponentially. That is to say, although precision values with high radius are higher than that with low radius~(aforementioned confusion situation), it's impractical in real applications. Hence we can conclude that RAMAP is more proper to evaluate hashing algorithms.

\section{Conclusion}
In this paper, we systematically analyze the problems of existing metrics for the first time and propose a novel evaluation metric, called radius aware mean average precision, to evaluate hash codes for bucket search. We propose two coding strategies to qualitatively show the problems of existing metrics. Experiments verify that our proposed metric can provide more proper evaluation for hashing.
\small
\bibliography{ref}
\bibliographystyle{abbrv}

\end{document}